\documentclass[12pt,preprint]{aastex}
\usepackage{graphics}

\def\gtorder{\mathrel{\raise.3ex\hbox{$>$}\mkern-14mu
              \lower0.6ex\hbox{$\sim$}}}
\def\ltorder{\mathrel{\raise.3ex\hbox{$<$}\mkern-14mu
              \lower0.6ex\hbox{$\sim$}}}

\shorttitle{Search for Comets Around HD209458}
\shortauthors{Mendelowitz, Ge, Mandell \& Li}

\begin{document}
\title{A Search for Sodium Absorption from Comets Around HD209458}
\author{Caylin Mendelowitz\altaffilmark{1}, Jian Ge\altaffilmark{1},
Avi M. Mandell\altaffilmark{1}, and Aigen Li\altaffilmark{2}}
\altaffiltext{1}{Pennsylvania State University, Department of
Astronomy \& Astrophysics, University Park, PA 16802, USA Email:
caylinm@astro.psu.edu, jian@astro.psu.edu, mandell@astro.psu.edu}
\altaffiltext{2}{Theoretical Astrophysics Program, Steward
Observatory, and Lunar \& Planetary Laboratory, University of Arizona,
Tucson, AZ 85721, USA Email: agli@lpl.arizona.edu}

\begin{abstract}
We monitored the planet-bearing solar-type star HD209458 for sodium
absorption in the region of the stellar NaI D1 line that would be
indicative of cometary activity in the system. We observed the star
using the HET HRS with high S/N and high spectral resolution for 6
nights over the course of two years, from July 2001 to July 2003. From
modelling we determine a 20\% likelihood of a detection, based on a
predicted number of comets similar to that of the solar system. We
find that our analytical method is able to recover a signal and our
S/N is sufficient to detect this feature in the spectral regions on
either side of the core of the D1 line, where it is most likely to
appear. No significant absorption was detected for any of the nights
based on a 3$\sigma$ detection limit. We derive upper limits on the
column density of sodium of $\lesssim 6 \times 10^{9}$ cm$^{-2}$ for a
signal in the region around the line core and $\lesssim 2 \times
10^{10}$ cm$^{-2}$ for a signal in the core of the photospheric D1
line. These numbers are consistent with the sodium released in a
single periodic comet in our own system, though higher S/N may be
necessary to uncover a signal in the core of the D1 line. Implications
for cometary activity in the HD209458 system are discussed.

\end{abstract}


\section{Introduction}
The Falling Evaporating Bodies (FEB) scenario \citep{Lagrange92} was
introduced as an explanation for time variable circumstellar
absorption features observed in the spectral lines of the A-type star
$\beta$ Pictoris \citep{Lagrange86,Beust00}. Similar features have
also been observed in other early-type main sequence stars and some
Herbig Ae/Be stars, which are considered $\beta$ Pictoris precursors
\citep{Welsh98,Grady97,Grinin96III}. According to the FEB hypothesis,
solid bodies with sizes typical of solar system comets, are evaporated
within the vicinity of the star producing absorption lines when
observed along the line of sight \citep{Ferlet87,Lagrange88}. The high
frequency of infalling comets is a phenomenon invoked in models of the
early solar system and corresponds to the clearing out stage in a
system's formation \citep{Welsh98,Beust98}. The mechanism behind the
infall of these bodies is thought to be from the gravitational
perturbation by at least one planet \citep{Beust00}. Monitoring known
planet-bearing solar analogs for comet activity would then give
insight into the evolution of an exosolar system and such a wide study
would investigate how comet populations evolve in time and with
stellar mass. Also, searches for comet disks and clouds orbiting other
stars offers a new method for inferring the presence of planetary
systems \citep{Stern93}.

Sungrazing comets and periodic comets are quite frequent in our own
system. Since the first light of the SOlar and Heliospheric
Observatory (SOHO) Large Angle and Spectrometric COronograph (LASCO)
on 1995 December 30 over 600 new sungrazing comets have been
discovered ($\sim$60/year) \citep{SOHOweb}. Also, the number of
catalogued periodic comets totals $\sim$900 with half having perihelia
$\le$ 1 AU \citep{Biswas00}. Comets evaporate a significant amount of
material as they approach perihelion in the form of a coma and
tails. For example, Halley's comet loses $\sim$0.001 of its mass every
perihelion passage ($\sim10^{11}$kg) \citep{Zeilik98}. Hale-Bopp is a
richer dust factory: in its 1997 apparition, it lost a total dust mass
of $\sim3 \times 10^{13}$ kg \citep{Jewitt99}. The absorption features
expected for many species known to be present in comets are quite weak
due to the small column densities. \citet{Cremoneseal97} discovered
the sodium tail in comet C/1995 O1 Hale-Bopp, which is a third type of
cometary tail consisting of neutral atoms. Sodium has also been
observed in the tail of some other bright comets and in the comae of
comets that were located at around 1 AU from the Sun (e.g., comet
1P/Halley) \citep{Wanatabe03}. Sodium is a good candidate for
extra-solar detection because it is detectable even when the column
density is low \citep{Cremoneseal97}, and sodium has already been
detected in absorption from gas in the $\beta$ Pictoris disk
\citep{Hobbs85}, in variable absorption in some Herbig Ae/Be stars
\citep{Sorelli96}, and in the atmosphere of the exosolar planet HD
209458b \citep{Charbonneau02}.
 
HD209458 is a solar-type star with a close-in Jovian-type planet. It
is oriented edge-on to our line of sight making it an ideal candidate
for absorption spectroscopy and the detection of orbiting bodies. This
has initiated several new exoplanet discoveries. It is the first
exosolar system for which repeated transits across the stellar disk
have been observed \citep{Charbonneau00}. This allowed
\citet{Charbonneau02} to make the first observation of an exosolar
planet atmosphere through the detection of neutral sodium
absorption. More recently, \citet{VidalMadjar03} detected atomic
hydrogen which is associated with the upper atmosphere of the
planet. Here we observe HD209458 for neutral sodium absorption from a
comet crossing the line of sight. We present spectra of HD209458 in
the vicinity of the Na D1 line (5895.9243 \AA). In $\S$2 we present
the data and describe the data reduction technique. We analyze the
data and remove the photospheric contribution of sodium. In $\S$3 we
use our simulation of a cometary sodium absorption feature to examine
its position in relation to the photospheric sodium line and its shape
based on expected values of the column density, in order to define a
search region. A synthetic spectrum is analyzed to check the
reliability of our method. In $\S$4 we examine the residuals, discuss
the results and derive upper limits for the column density of sodium
on each night. We use modelling to determine the likelihood of
detecting a comet and conclude with possible explanations for the
non-detection and a discussion of comet activity in the system.

\section{Observations \& Analysis}
Observations of HD209458 were made with the High Resolution
Spectrograph (HRS) at the 9.2 meter Hobby-Eberly Telescope (HET)
\citep{Tull98}. The HRS is a fiber-coupled echelle spectrograph, using
an R-4 echelle mosaic with cross-dispersing gratings. The camera
images onto a mosaic of two 2K x 4K CCDs with 15 $\mu$m pixels and a
gap between them that spans $\sim$72 pixels. The spectrograph was used
in high-dispersion mode, with an average resolving power of $R =
\lambda/\Delta\lambda \sim 115,000$ measured from thorium lines,
corresponding to a resolution element of $\sim$0.049 \AA\ in the
region of interest. The spectral coverage for the cross disperser
setting used is between 4076 \AA\ and 7838 \AA. Observation dates and
parameters are listed in Table \ref{observations}. The star HD12235
was also observed to serve as a reference system. We observed for two
nights on 2002 August 28 and 31, using the same instrumental settings
as HD209458. Stellar parameters are listed in Table \ref{params}.

\begin{figure}
\begin{center}
\includegraphics[scale=0.8]{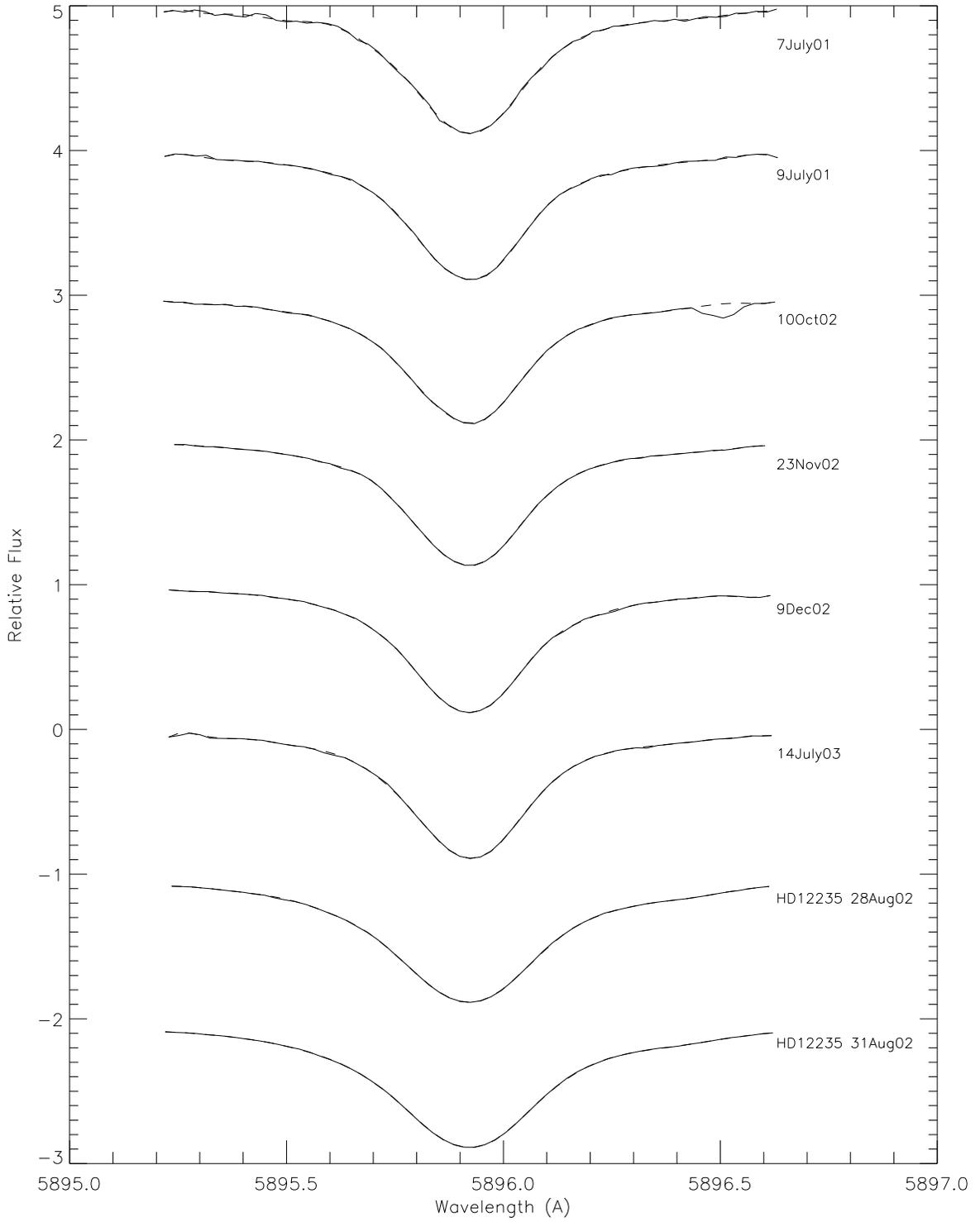}
\caption{\label{plotone} HD209458 on all 6 nights and HD12235. The fit to
the D1 line is overplotted with a dashed line. The spectra have been
shifted to the rest frame of the system.}
\end{center}
\end{figure}

Images typically had an integration time of 5 to 25 minutes depending
on brightness, and a total integration time between 1 and 2 hours. A
Thorium-Argon lamp was observed before and after each sequence of
exposures. Due to sodium emission lines in the flatfield illumination
lamp, it was necessary to use extremely high S/N observations of
rapidly rotating B and A stars as flat fields. These calibration stars
were observed as near in time to the program stars as possible. The
data were reduced using the publically available IRAF software
package. Continuum normalization was performed by selecting continuum
regions in each image without obvious absorption lines or bad pixels
and fitting a higher order polynomial to these sections. Images were
then combined with appropriate rejection algorithms. Wavelength
calibration of the spectra was obtained using the Thorium-Argon lamp
comparison spectra for each data set. The 2002 December 9 data was
missing a calibration star so the lamp was used for flatfielding, with
the sodium emission lines removed using a polynomial fit to obtain a
featureless spectrum. The S/N per pixel in the continuum for each
image is listed in Table \ref{upplim}. Figure \ref{plotone} shows the
reduced spectra for HD209458 for each of the 6 nights. The spectra
have been shifted to the rest frame of HD209458.

The D1 line was used for analysis since the D2 line is heavily
contaminated by telluric lines \citep{Ge98}. The broad photospheric
sodium line was removed using a polynomial fit (see Figure
\ref{plotone}), including multiple rejection iterations and the
normalized spectrum was searched for possible narrow absorption
features up to 0.7 \AA\ from the photospheric line center. Several
absorption features above 3$\sigma$ were attributed to telluric
features that were removed imperfectly due to changes in observing
conditions between exposures of HD209458 and the calibration
star. This is the case for the 2002 October 10 spectrum where there is
a significant absorption feature around 5896.5 \AA; the feature is
also present in the calibration star used for flatfielding. For 2001
July 7 observing conditions were deteriorated by the fourth exposure,
which is the cause of the absorption feature near 5895.86 \AA. No
absorption features greater than 3$\sigma$ due to cometary activity
were identified.

\section{Modelling}
Simulations were run to produce estimates of the position and shape of
a sodium absorption feature due to solar-like comets in the HD209458
system. Constraints on spatial and chemical characteristics of
cometary bodies were taken from observations of short and long period
comets in our own system, mainly the comets Halley and Hale-Bopp
\citep{Swamy97,Zeilik98,Cremoneseal97}. Although sungrazing comets
also display tail and coma features, they are much smaller than
periodic comets, only a few meters to a few tens of meters
\citep{Iseli02} (whereas periodic comets are a few kilometers to
hundreds of kilometers), and therefore will not release a significant
amount of dust. The position of the feature will depend on the
velocity of the comet when detected; as a comet moves within $\sim$3
AU the cometary nucleus begins to display coma formation followed by
the extension of gas and ion tails. Full display of the comet tail
occurs between 1.0 and 1.5 AU, \citep{Biswas00} with sodium extending
beyond 10$^{5}$ km from the nucleus \citep{Oppenheimer80}. In
Hale-Bopp the length of the sodium tail was measured to be 3$\times
10^{7}$ km \citep{Cremonese97}. In order to obtain the maximum column
density along the line of sight, a theoretical comet velocity was
calculated specifically for an orientation aligned with the sodium
tail, which points roughly anti-sunward. Since the radius and
luminosity of HD209458 are larger than for the Sun, the orbital
distance for detecting the sodium tail was set to within 3 AU of the
star.

\begin{figure}
\includegraphics[scale=0.5]{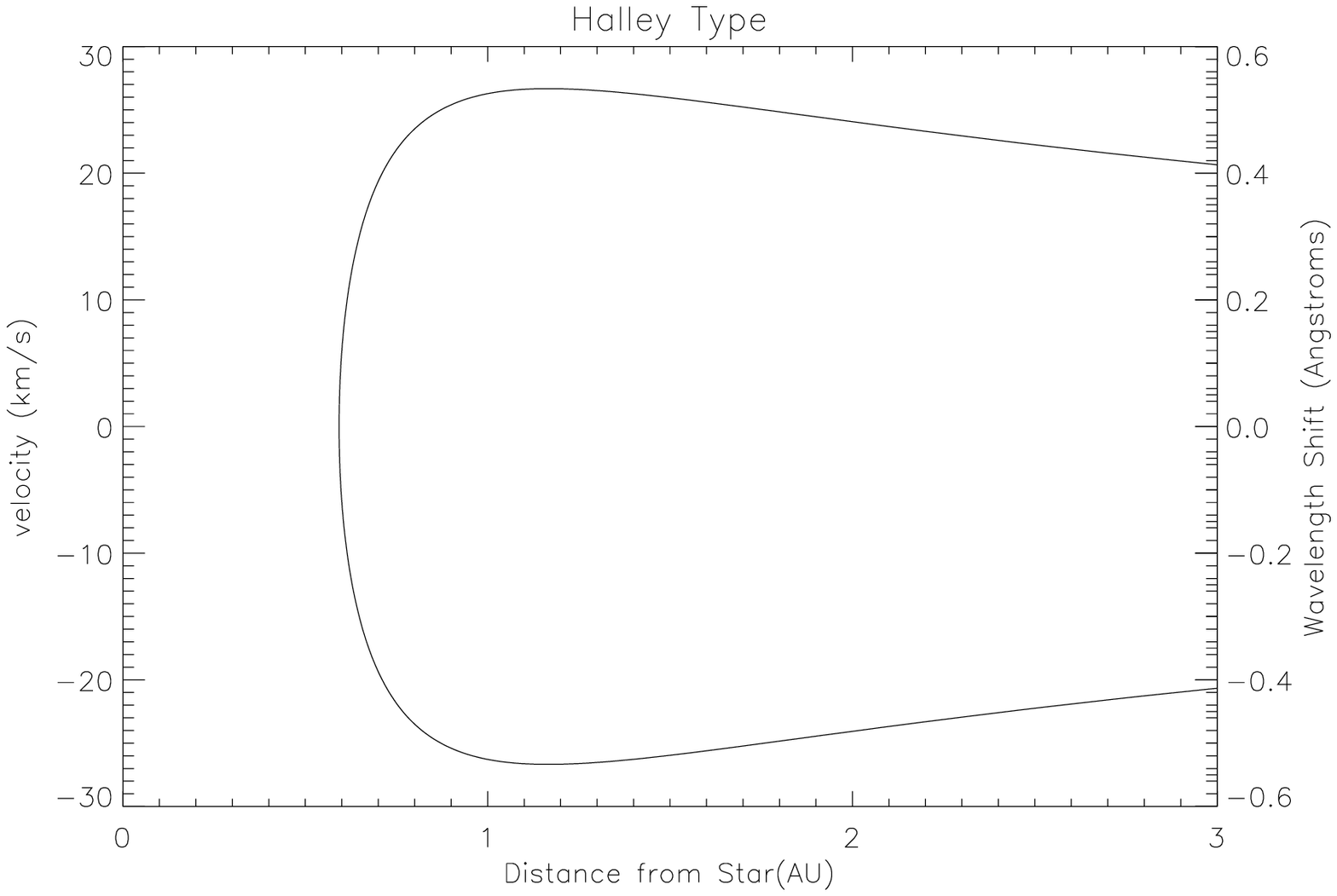} \includegraphics[scale=0.5]{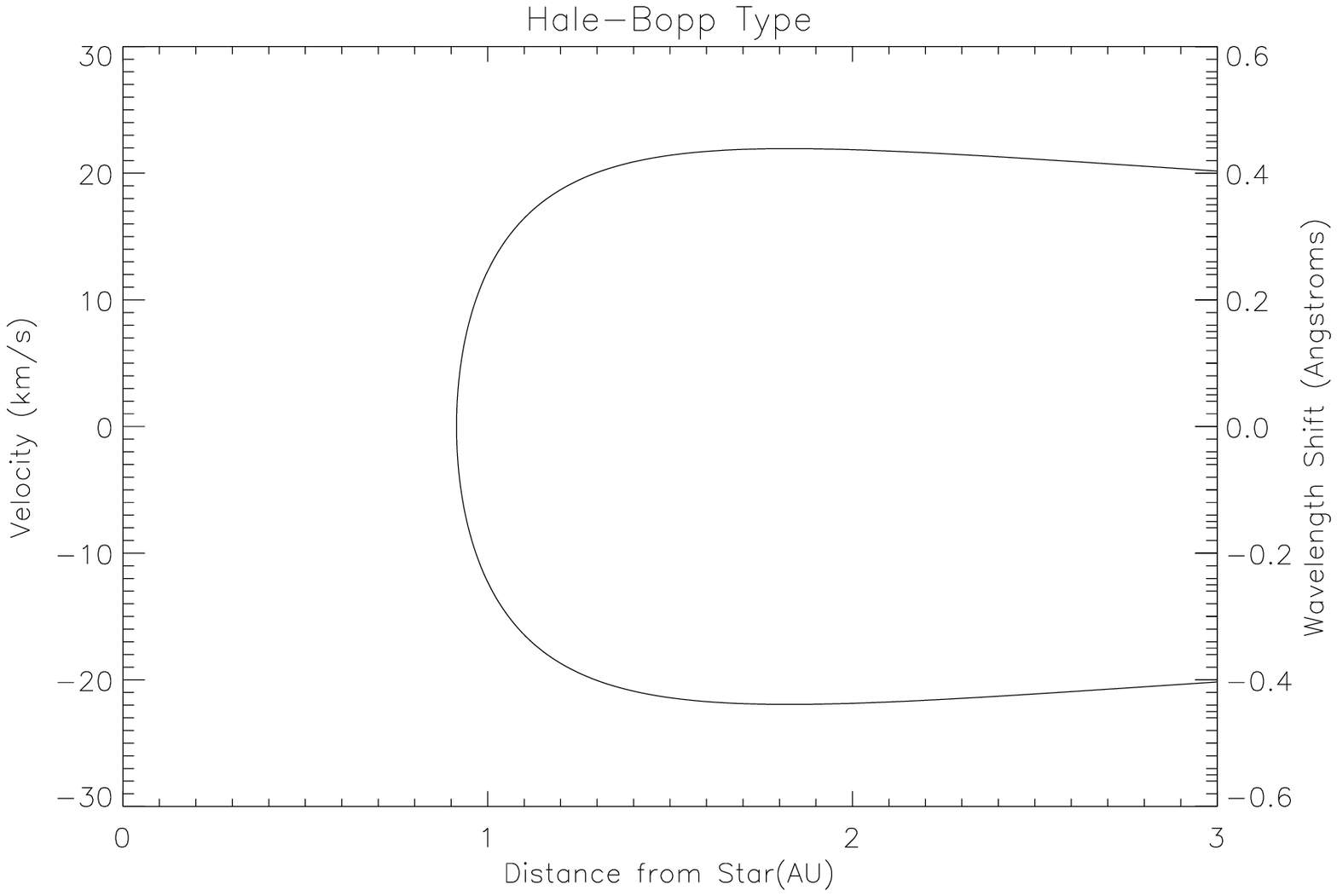}
\caption{\label{plottwo} Wavelength and corresponding doppler shift due
to the tail velocity for (Left) Short period comets and (Right) Long
period comets. The above plots show that the greatest wavelength shift
expected would be 0.53 \AA\ from the photospheric line center.}
\end{figure}

The wavelength shift due to the orbital motion of the comet with
respect to the parent star is described by the radial velocity of the
comet:
\begin{equation}
v_{r}=(2a/P)(e\sin\theta)(1-e^{2})^{-1/2}
\end{equation}
where $e$ is the eccentricity, $a$ is the semi-major axis, $P$ is the
period and $\theta$ is measured so that $\theta$=0 is at
perihelion. The distance from the star is given by:
\begin{equation}
r = \frac{a(1-e^{2})} {e\cos\theta+1}
\end{equation}
The range of velocity shifts for a spectral feature have been
calculated using the parameters of the short-period (SP) comet Halley
($e$=0.967, a=18 AU, P=76 years, perihelion distance q=0.587 AU) and
the long-period (LP) comet Hale-Bopp ($e$=0.9951, a=187 AU, P=2550
years, q=0.914 AU). Figure \ref{plottwo} shows the wavelength and
corresponding doppler shift for both types of comet. The velocity and
thus the wavelength shift for Halley peaks at 1.17 AU, with a maximum
shift of 0.53 \AA\ from the photospheric line center. For Hale-Bopp
the maximum velocity is reached at 1.83 AU, which results in a maximum
wavelength shift of 0.43 \AA. The majority of the orbit, from
$\sim$1-3 AU, for both types of comet will produce a wavelength shift
of $\sim$0.4 \AA. Since the D1 line is $\sim$0.6 \AA\ wide, a feature
will most likely be in the wing or continuum region. We also take into
consideration the periodic comets with perihelia less than those of
Halley and Hale-Bopp ($\sim$220 comets \citep{Biswas00}), which would
produce a greater wavelength shift. Our search for Na absorption will
be $\pm$0.7 \AA\ around the stellar Na D1 line.

\begin{figure}
\includegraphics[scale=0.5]{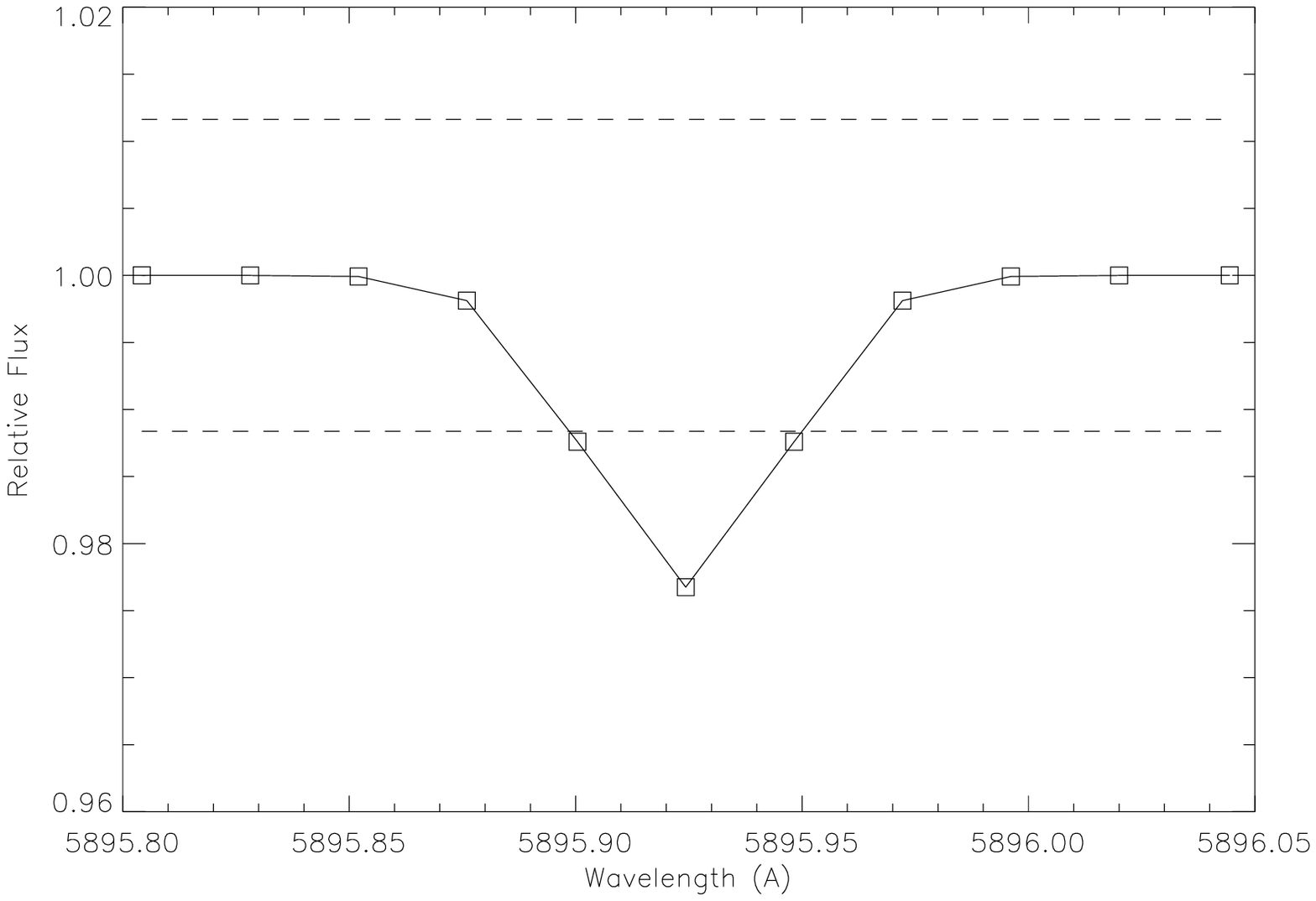}
\includegraphics[scale=0.5]{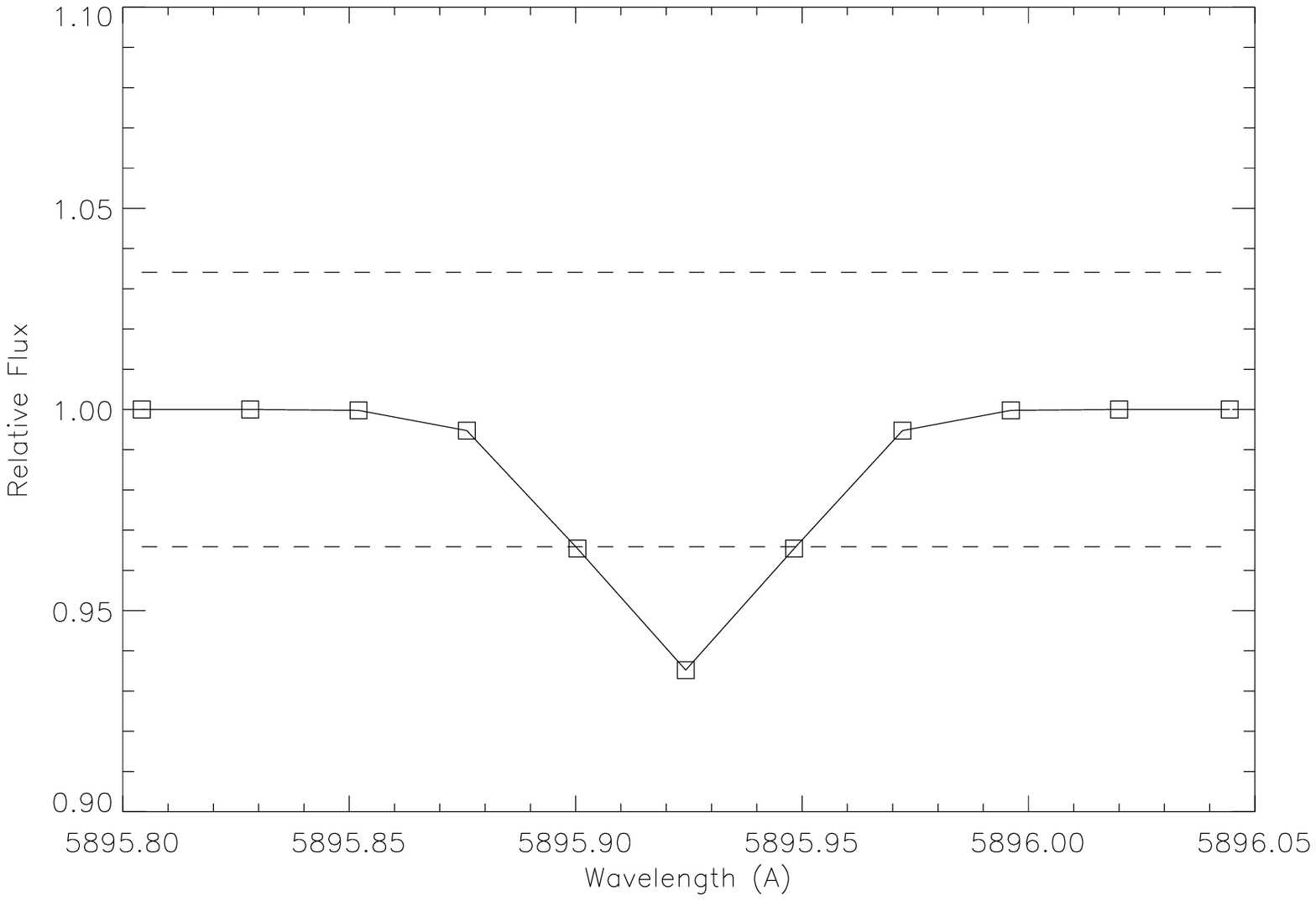}\\ 
\caption{\label{plotthree} Theoretical Sodium Absorption Line. (Left)
Column density of 5.8 $\times 10^{9}$ cm$^{-2}$ based on upper limit
of best S/N in the continuum ($\sim$260 for 2002 Dec 09). Dashed lines
represent 3$\sigma$. (Right) Column density of 1.7 $\times 10^{10}$
cm$^{-2}$ based on upper limit of best S/N in the core of the D1 line
($\sim$90 for 2002 Dec 09). Dashed lines represent 3$\sigma$.}
\end{figure}

The depth and shape of the feature depends on the column density and
temperature of the absorbing material. A model of the signal was
created using a Voigt absorption profile, which was then convolved
with the instrumental spectral resolution. Figure \ref{plotthree}
shows model absorption lines based on the core and continuum upper
limits derived for our best S/N data, 2002 December 9, in
$\S$4.1. Estimates of the column density of sodium from evaporating
comets in other systems are on the order of $10^{10}$ cm$^{-2}$ from
HR10 (FEB) \citep{Welsh98} and from $\beta$ Pictoris (disk)
\citep{Hobbs85}. The thermal broadening can be calculated by assuming
a gas temperature for the comet coma and tail. The estimated
temperature for the sodium tail of a comet, from solar system
observations, is 150K \citep{Swamy97}, and atomic parameters for the
sodium D1 line were taken from \citet{Morton91}. The inherent Doppler
width at 150K corresponds to only 0.03 \AA, and therefore a single
radial velocity absorption component would be spectrally unresolved,
with the instrumental FWHM of 0.049 \AA\ of the HRS. The equivalent
width of a sodium absorption line with a column density of 10$^{10}$
cm$^{-2}$ is $\sim$1 m\AA.

\subsection{Synthetic Detection: Check on the Analysis}
\begin{figure}
\begin{center}
\includegraphics[scale=0.9]{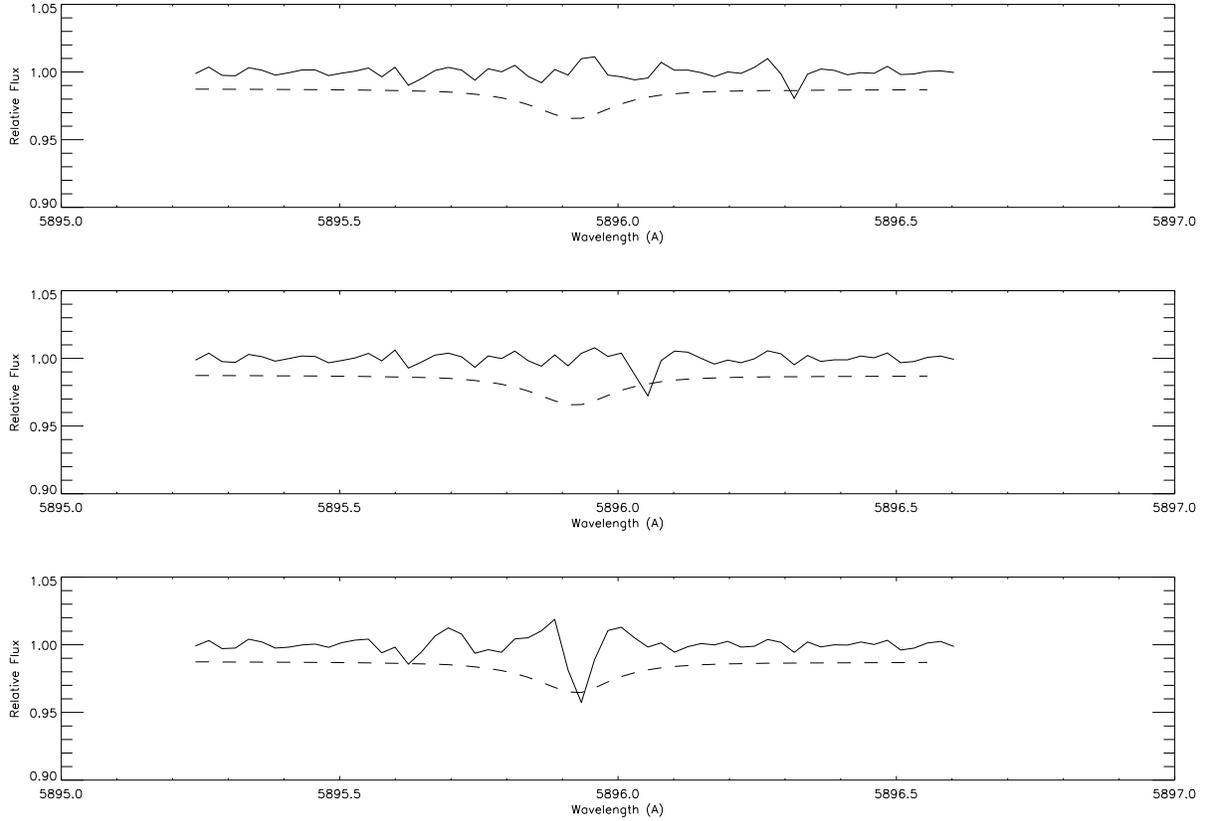}
\caption{\label{plotfour} Residual from polynomial fit to the 2002
November 23 stellar sodium D1 profile. The absorption feature is a
sodium synthetic feature in the continuum region with N(NaI)=6.4
$\times 10^{9}$ cm$^{-2}$ (Top) in the wing with N(NaI)=9.3 $\times
10^{9}$ cm$^{-2}$ (Middle) and in the core with N(NaI)=1.7 $\times
10^{10}$ cm$^{-2}$ (Bottom). Column densities from 3$\sigma$ upper
limits listed in Table 3. Dashed lines represent 3$\sigma$.}
\end{center}
\end{figure}

As a check of our analytical method we input our synthetic absorption
features into the observed spectrum and recover the absorption through
fitting the broad stellar sodium D1 line profile. We use the data from
2002 November 23 since telluric features have been successfully
removed and the S/N is the highest (They were not removed in 2002
December 9). Since the detection limit is lower in the continuum than
in the core of the line, we place the feature in the continuum
(S/N=237) with a column density of 6.4$\times 10^{9}$ cm$^{-2}$ and in
the core (S/N=88) with a column density of 1.7$\times 10^{10}$
cm$^{-2}$ derived in $\S$4.2 as the 3$\sigma$ upper limits for this
date. We also place a feature in the wing of the line with a column
density of 9.3 $\times 10^{9}$ cm$^{-2}$ based on a calculated S/N of
162. The feature was added at the stage just before the removal of the
photospheric D1 line. Our method of polynomial fitting was able to
recover the features at the core, wing and also continuum region. The
residuals and recovered absorption features are shown in Figure
\ref{plotfour}. The equivalent width of the recovered features are
1.6$\pm$0.6, 1.0$\pm$0.3, and 0.62$\pm$0.2 m\AA, for the core, wing,
and continuum respectively. These are all consistent with the input
values of 1.7, 0.9, and 0.6 m\AA, for the core, wing, and continuum
respectively. This demonstrates proof that the stellar D1 line profile
fitting technique can recover narrow sodium features from comet
events.

\section{Results and Discussion}
\subsection{Upper Limits on Sodium Column Density by Comets}
\begin{figure}
\includegraphics[scale=0.5]{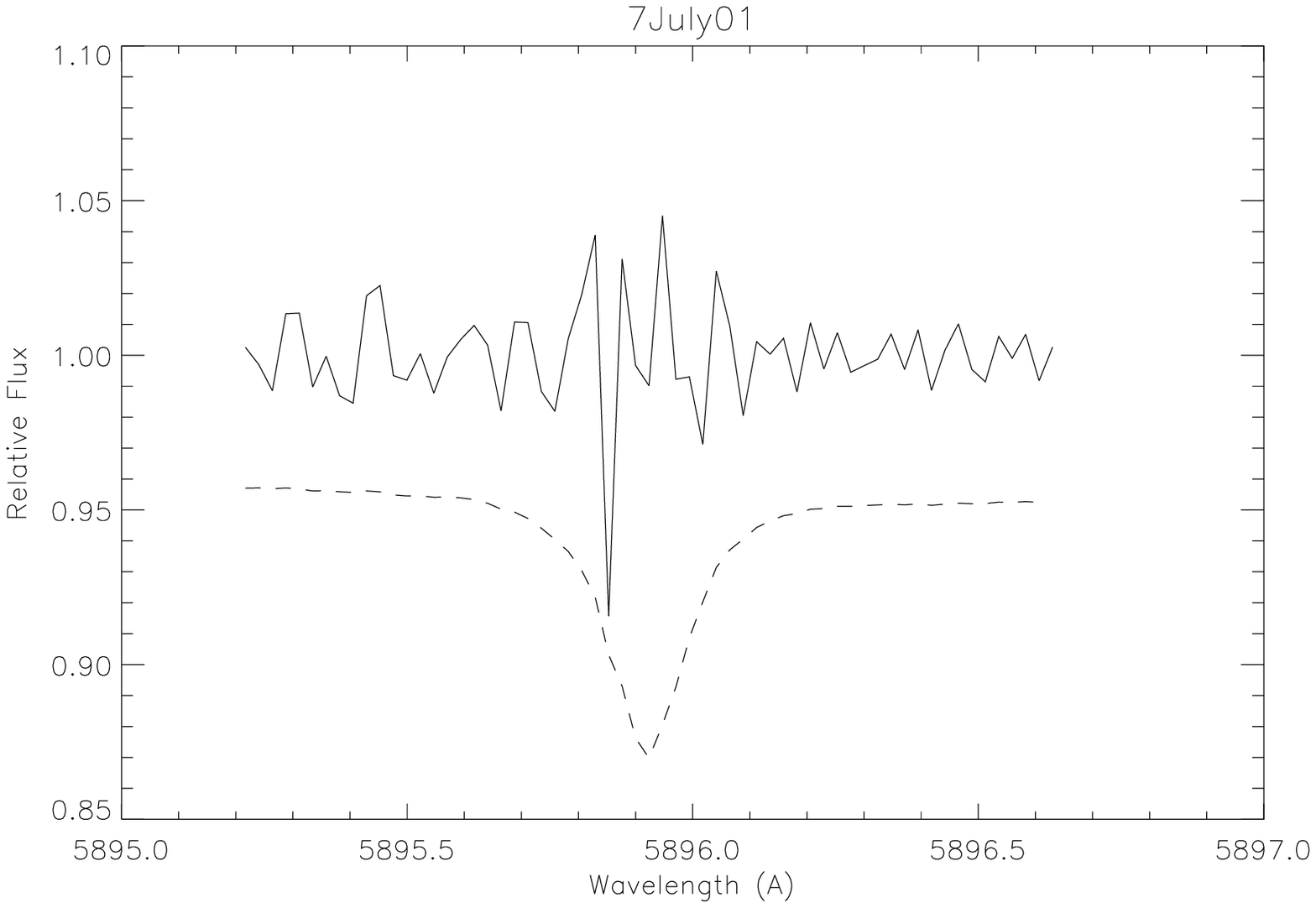}
\includegraphics[scale=0.5]{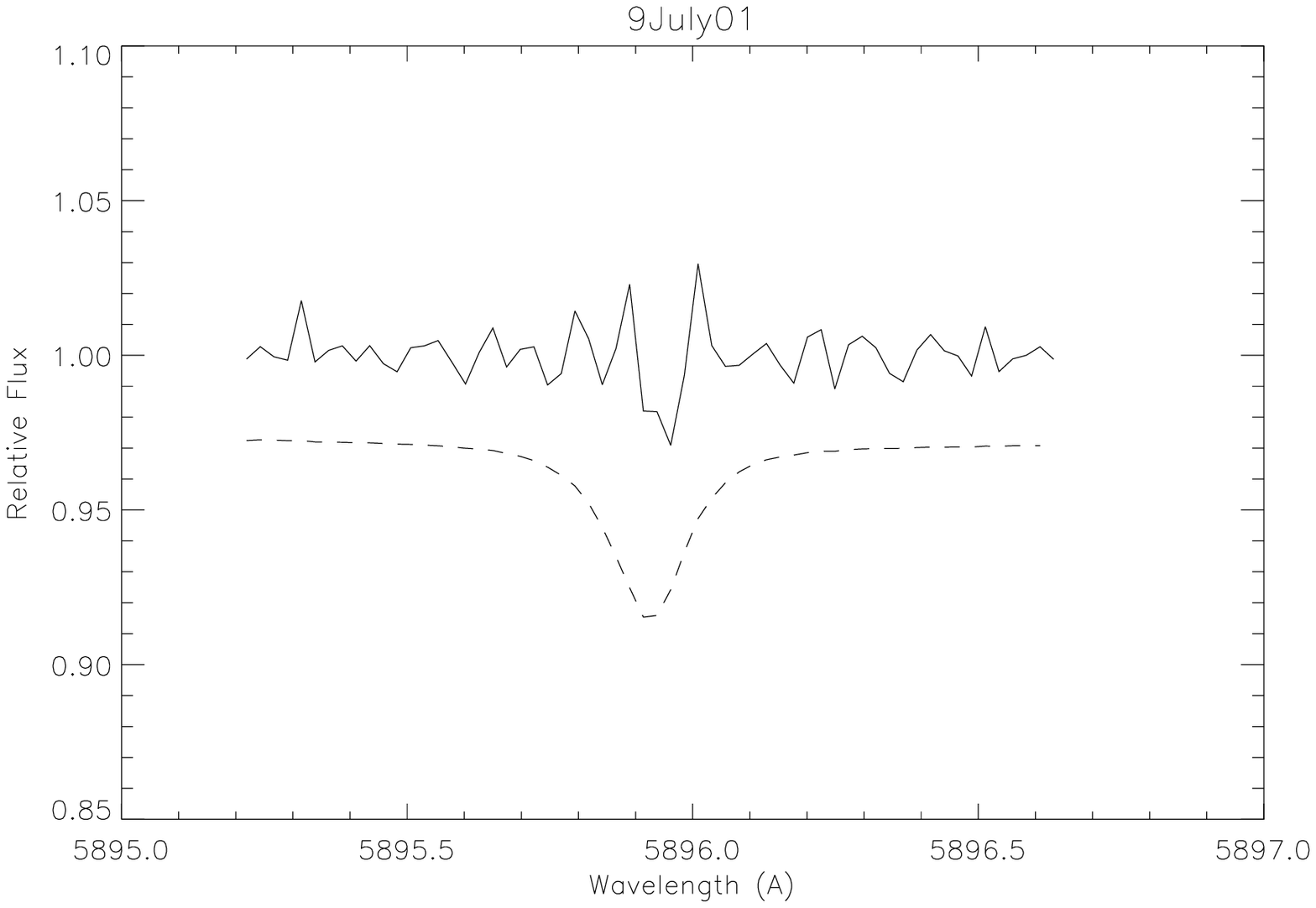}
\includegraphics[scale=0.5]{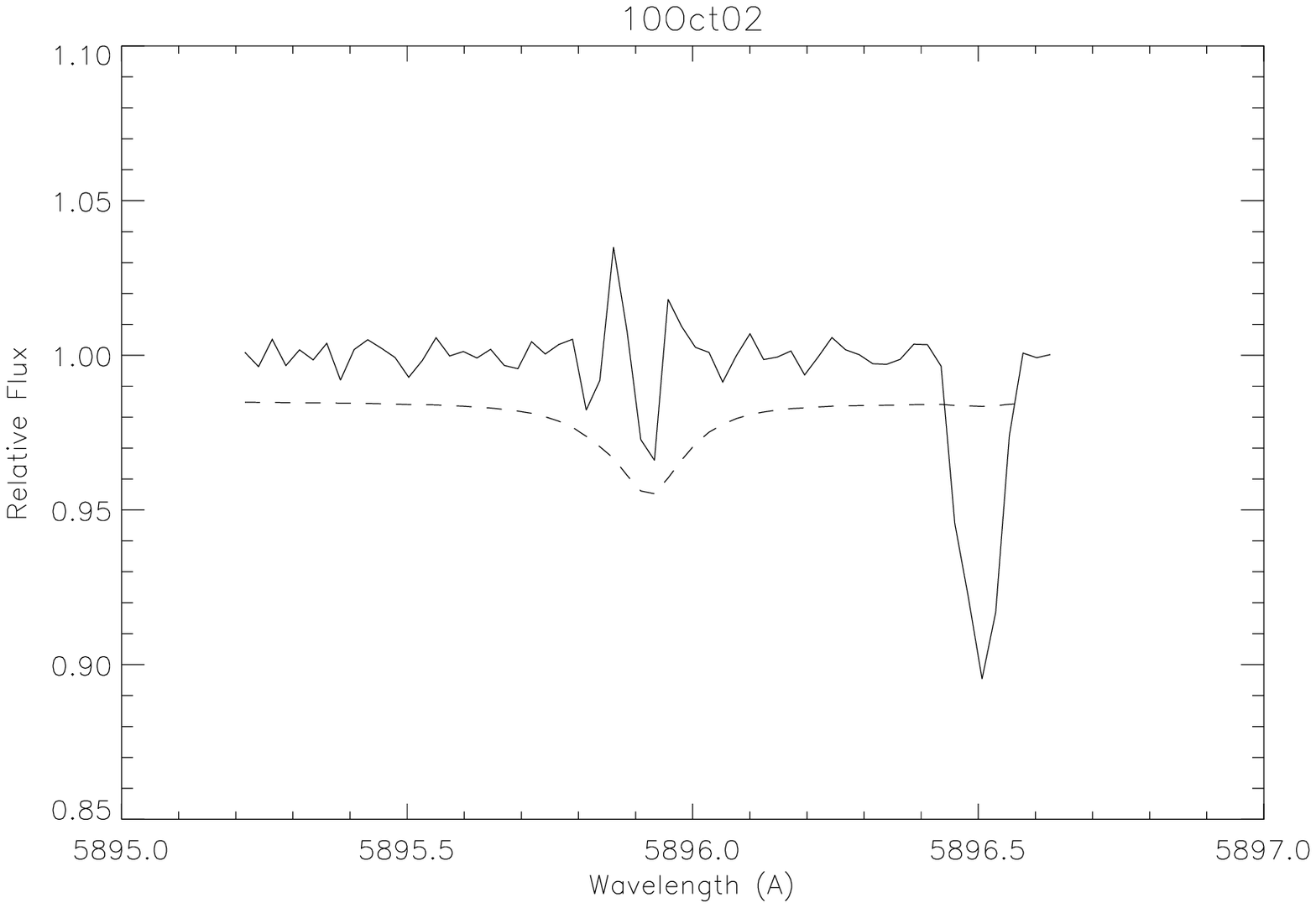}
\includegraphics[scale=0.5]{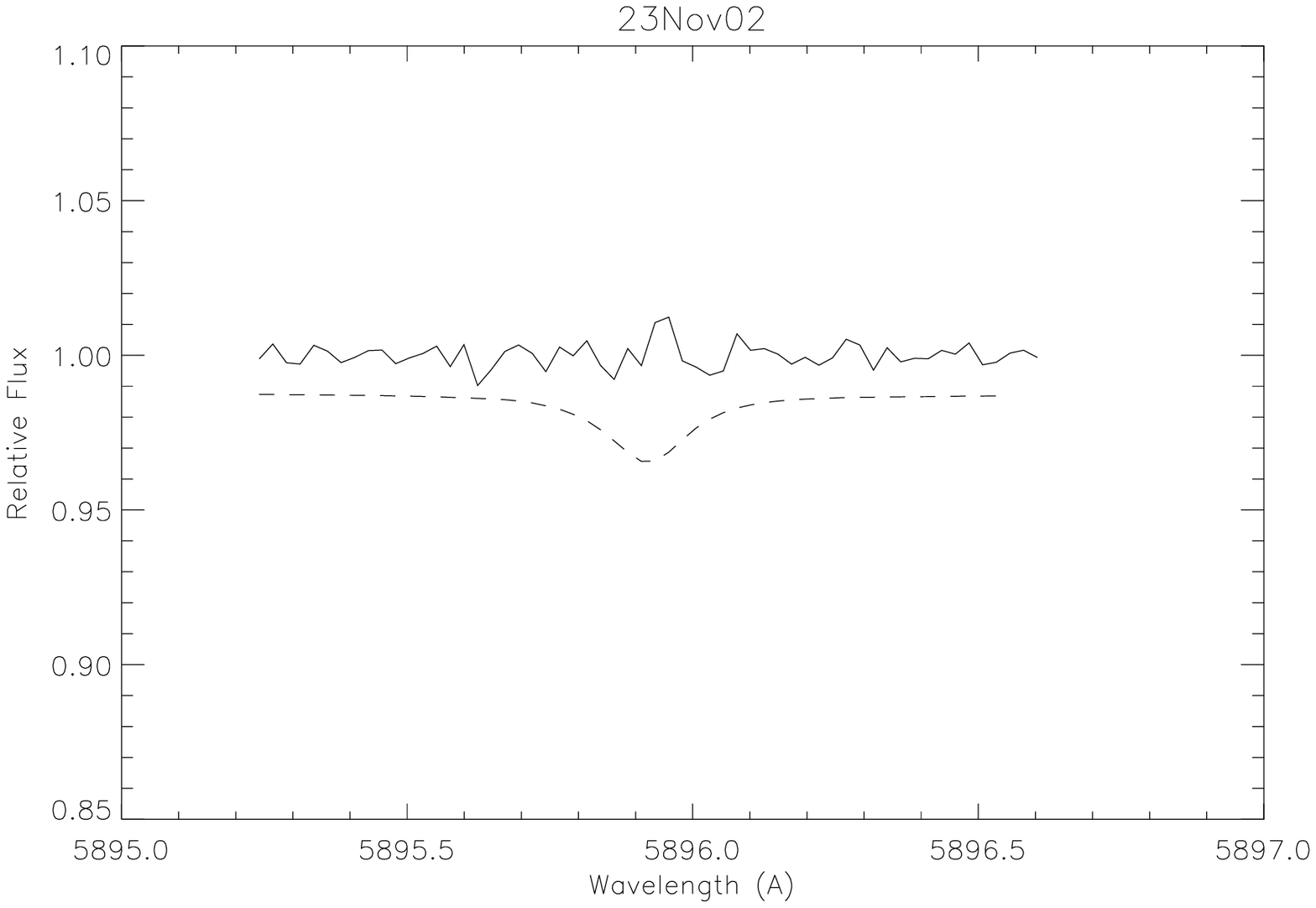}
\includegraphics[scale=0.5]{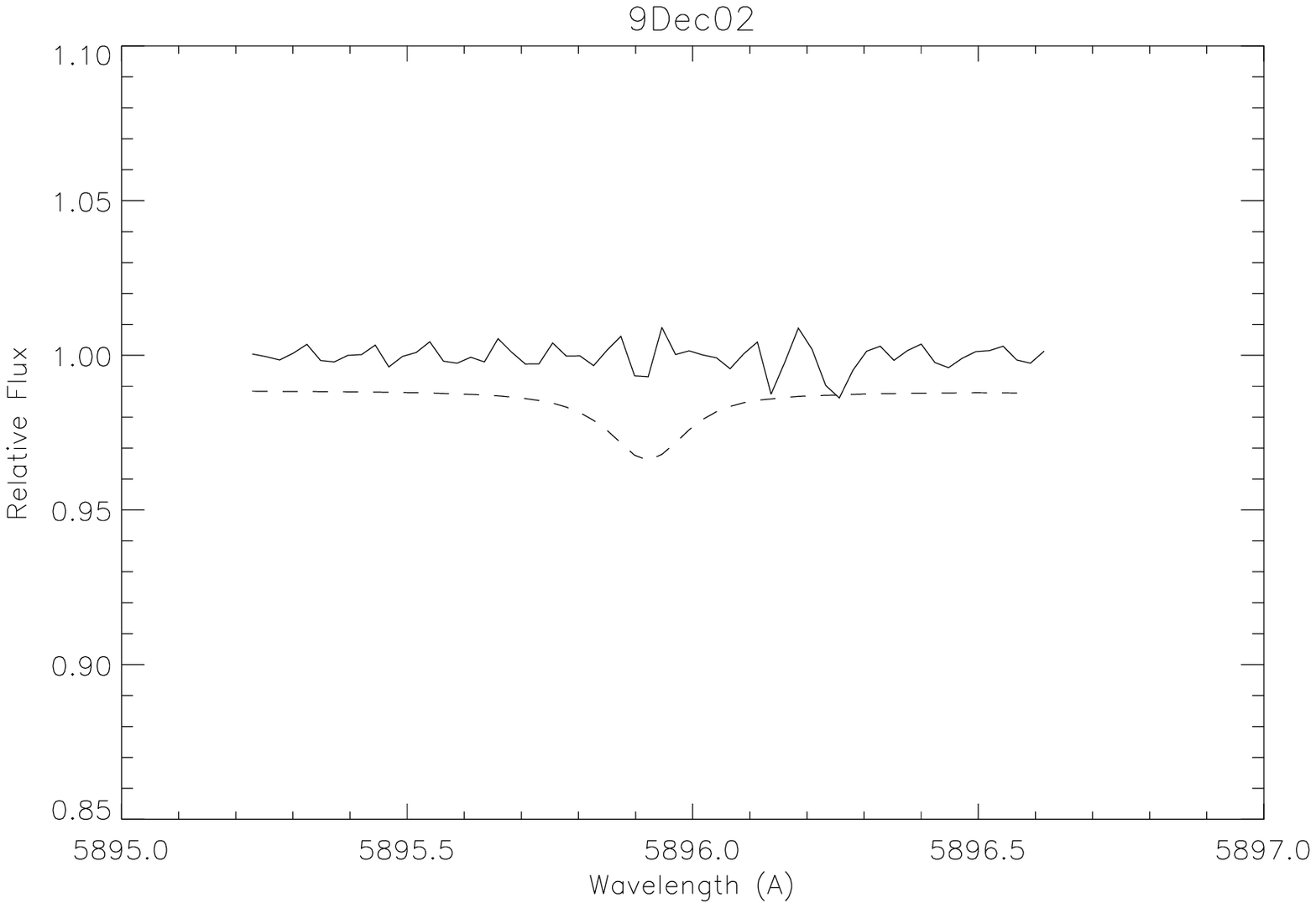}
\includegraphics[scale=0.5]{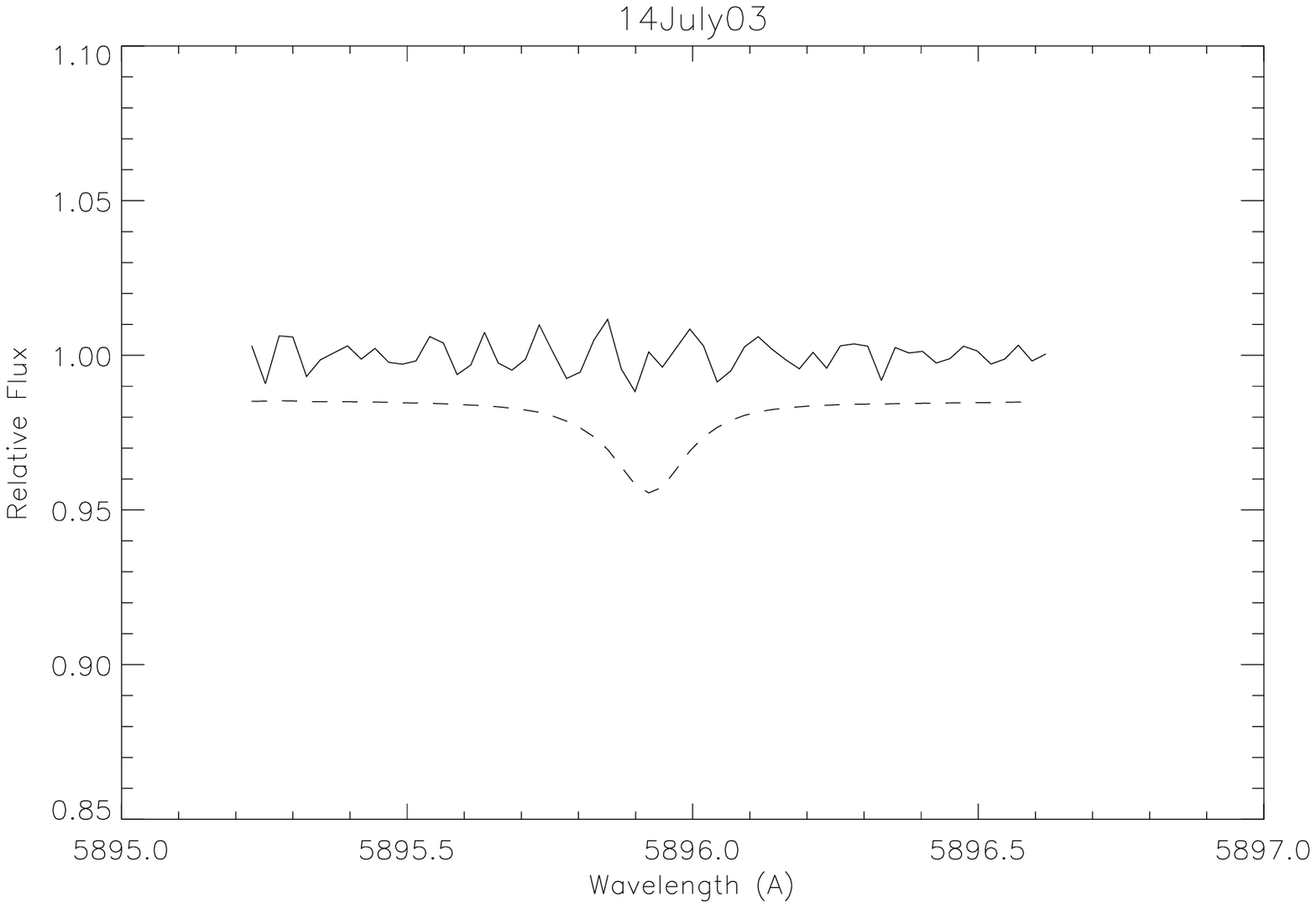}
\caption{\label{plotfive} Residual from polynomial fit to the
photospheric D1 profile. Dashed line represents 3$\sigma$. In 2002
October 10 the absorption feature at 5896.5 \AA\ is from a telluric
line that was poorly removed.}
\end{figure}

\begin{figure}
\includegraphics[scale=0.5]{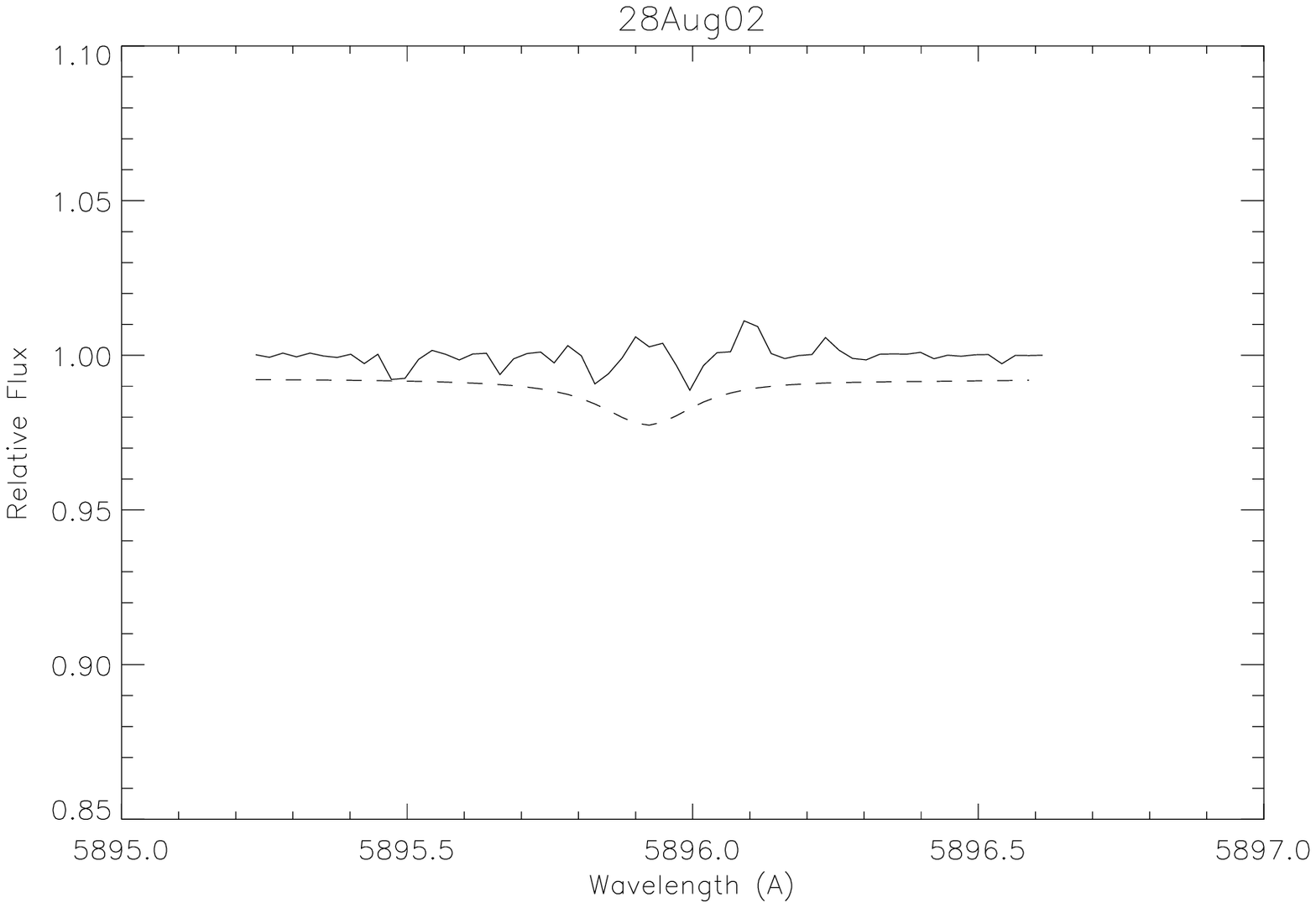}
\includegraphics[scale=0.5]{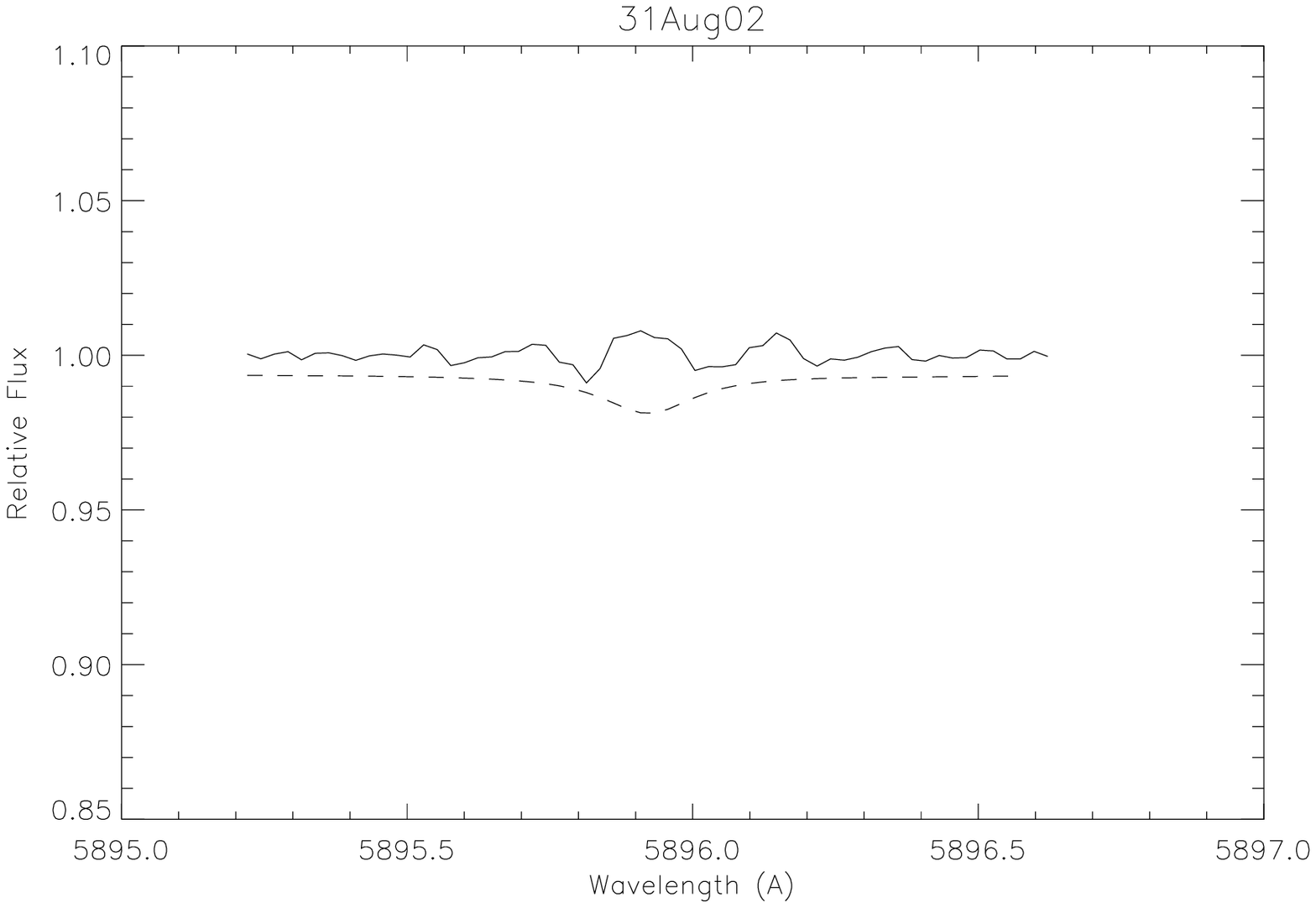} 
\caption{\label{plotsix}Reference star HD12235. Residual from
polynomial fit to the photospheric D1 line. Dashed line represents
3$\sigma$.}
\end{figure}

The residual after the photospheric sodium line is removed are shown
in Figure \ref{plotfive}. No signal was detected in any of the spectra,
but upper limits were determined based on a 3$\sigma$ detection
limit. The 1$\sigma$ upper limit to the equivalent width for the
undetected feature was determined by:
\begin{equation}
W_{\lambda} = \int \frac{F_{c} - F_{\lambda}}{F_{c}} d\lambda
\approx \Delta\lambda \times \sigma \\ 
\end{equation}
where $F_{c}$ is the normalized flux in the continuum, $F_{\lambda}$
is the normalized absorption line, $\Delta\lambda$ is the spectral
resolution element 0.049 \AA, and $\sigma$ is:
\begin{equation}
\sigma(\lambda) = \frac{1} {F_{\lambda}I_{c}(\lambda)} 
\{F_{\lambda}I_{c}(\lambda) + p\varepsilon^{2}\}^{1/2} \approx \frac{1}{S/N}
\end{equation}
$I_{c}(\lambda)$ = $N_{c}$ $\times$ gain, where $N_{c}$ is the number
of counts in the continuum and gain=0.6320 \emph{e}$^{-}$ ADU$^{-1}$,
$\varepsilon$ is the readnoise 2.70 \emph{e}$^{-}$, and $p$ is the
pixel number in the cross-slit direction used for sampling a
spectrum. It is appropriate to assume that the absorption line is in
the linear portion of the curve of growth, where the column density is
proportional to the equivalent width. The integrated number of sodium
atoms per square centimeter in the line of sight is then given by:
\begin{equation}
N = \frac{mc^{2}}{\pi e^{2} f} \int_{0}^{\infty} \frac{F_{c}(v) -
F(v)}{F_{c}(v)} dv = 1.130 \times 10^{20} \frac{W_{\lambda}}
{f\lambda^2} cm^{-2},\\
\end{equation}
where $f$ is the oscillator strength ($f$=0.318 for 5895.9243
\citep{Morton91}, and $W_{\lambda}$ and $\lambda$ are in units of
Angstroms. The upper limits for each date are given in Table
\ref{upplim}. In the continuum region, the column density for each
night ranges from $\sim6\times 10^{9}-2\times 10^{10}$ cm$^{-2}$,
whereas in the region of the D1 line core upper limits range from
$\sim2\times 10^{10}-7\times 10^{10}$ cm$^{-2}$. This appears to be
consistent with the solar comet sodium column density (e.g., sodium
absorption from an occultation of the solar system comet
Giacobini-Zinner resulted in upper limits of N(Na{\footnotesize I}) $<
2 \times 10^{11}$ cm$^{-2}$, based on solar abundance ratios for
K{\footnotesize I} \citep{Schempp89}). For comparison, we also show
the residual for the reference star HD12235 in Figure
\ref{plotsix}. This star was reduced in exactly the same manner as
HD209458 and the stellar sodium D1 line was fit and removed using the
same methods as for HD209458. As expected, no features were present in
the spectra on either night. The 3$\sigma$ upper limits are 1.1
$\times 10^{10}$ cm$^{-2}$ (2002 August 28), and 9.4 $\times 10^{9}$
cm$^{-2}$ (2002 August 31), which are in agreement with our results
for HD209458.

\subsection{Detection Probability}
Since no signal was detected we examine the likelihood of observing a
comet crossing based on the amount of time the comet spends within an
observable window. We only consider the orbit within 3 AU since comet
features will not be significant beyond this distance from the
star. The amount of time the comet spends in front of the star is
determined by the diameter of the star, D$_{sun}$, and the tangential
velocity, v$_{tan}$. This time will vary depending on the orientation
of the orbit to our line of sight. The radius of HD209458 is
1.18R$\odot$ $\approx 8.21 \times 10^{5}$ km, which is comparable to
the width of the sodium tail, 8$\times 10^{5}$ km
\citep{NASAHaleBopp}. For the total time in front of the star within a
year for any comet in the system we multiply by the average number of
periodic comets per year $\sim$25, $\sim$10 of which are SP comets
(Halley type) and $\sim$15 of which are LP comets (Hale-Bopp type)
\citep{Swamy97}. We double this since our observations are over the
course of two years. The time that any SP comet spends in front of the
star over two years ranges between 7 days and 35 days for v$_{tan}$
varying from 11 km s$^{-1}$ to 54 km s$^{-1}$, with an average time of
in front of the star of 10 days. For LP comets the time ranges between
13 days and 42 days for v$_{tan}$ varying from 14 km s$^{-1}$ to 44 km
s$^{-1}$, with an average time of 18 days in front of the star. Since
our observing time is small compared to the time comets spend in front
of the star and observing a comet can be considered a random process,
the probability can be efficiently described by a Poisson distribution
$P(t,\tau)=1-e^{-\mu}$ where $e^{-\mu}$ is the probability of
observing zero events over the total combined observing window, and
$\mu=(\tau N)$ is the average number of events for the combined
observations. $\tau$ is defined as the average number of comets
observable at any time, which can be calculated by multiplying the
total amount of time that a single comet would spend in the line of
sight by the estimated number of comets per year and then dividing by
the the total time (one year). N is the total number of chances to see
a comet, defined as the total number of observing sessions. An
observing session is defined as the period of time needed to reach a
S/N necessary to detect sodium absorption. This calculation assumes
that the observable windows for the different comets do not overlap, a
reasonable assumption considering the estimated number of comets and
amount of time each one spends in front of the star. Since the
estimated number of comets per year can be considered to be constant
over our monitoring period, the chances of observing a comet are
insensitive to the elapsed time between observations, and depend only
on the average number of comets per year in the system and the total
number of observing sessions.

\begin{figure}
\includegraphics[scale=0.5]{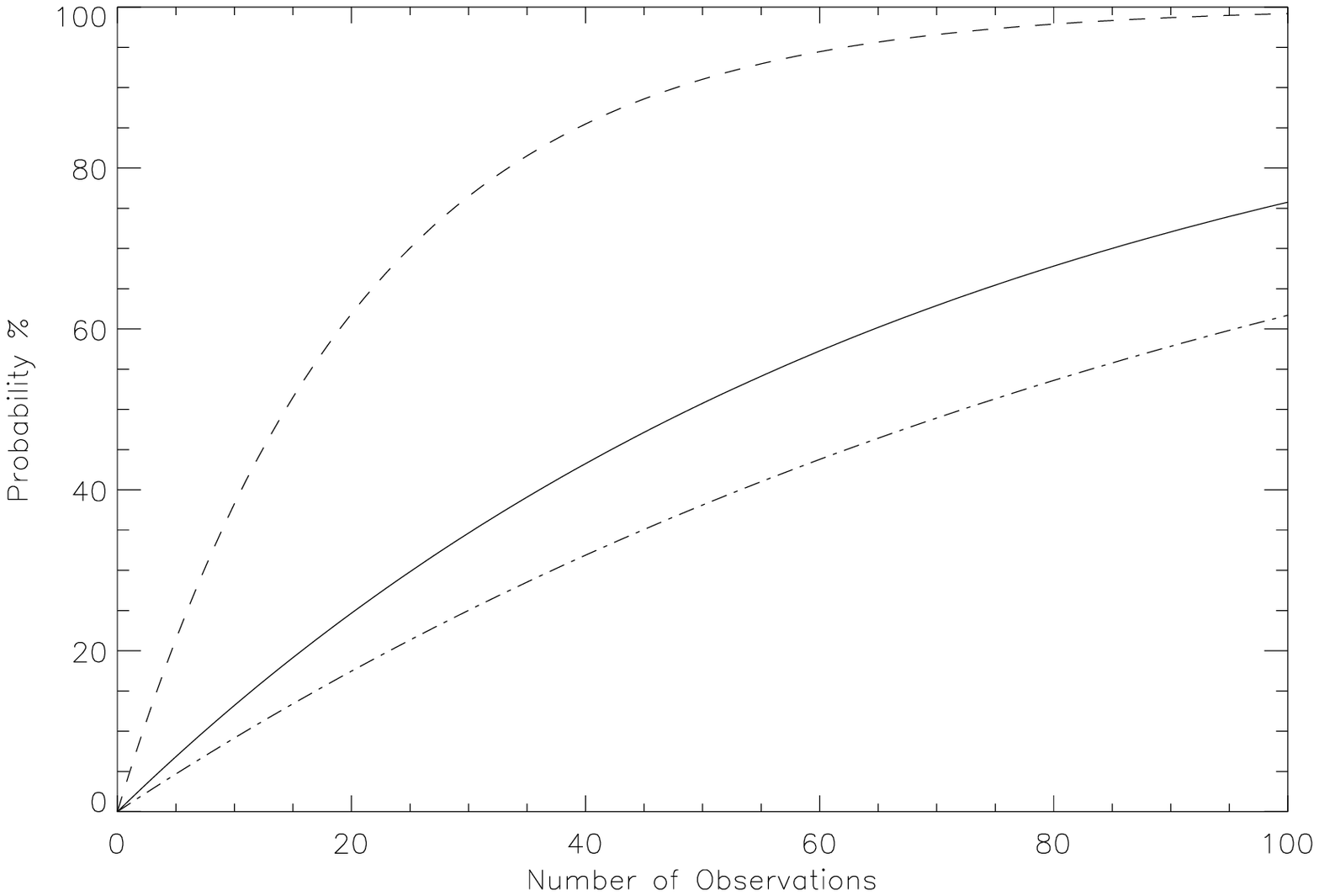}
\includegraphics[scale=0.5]{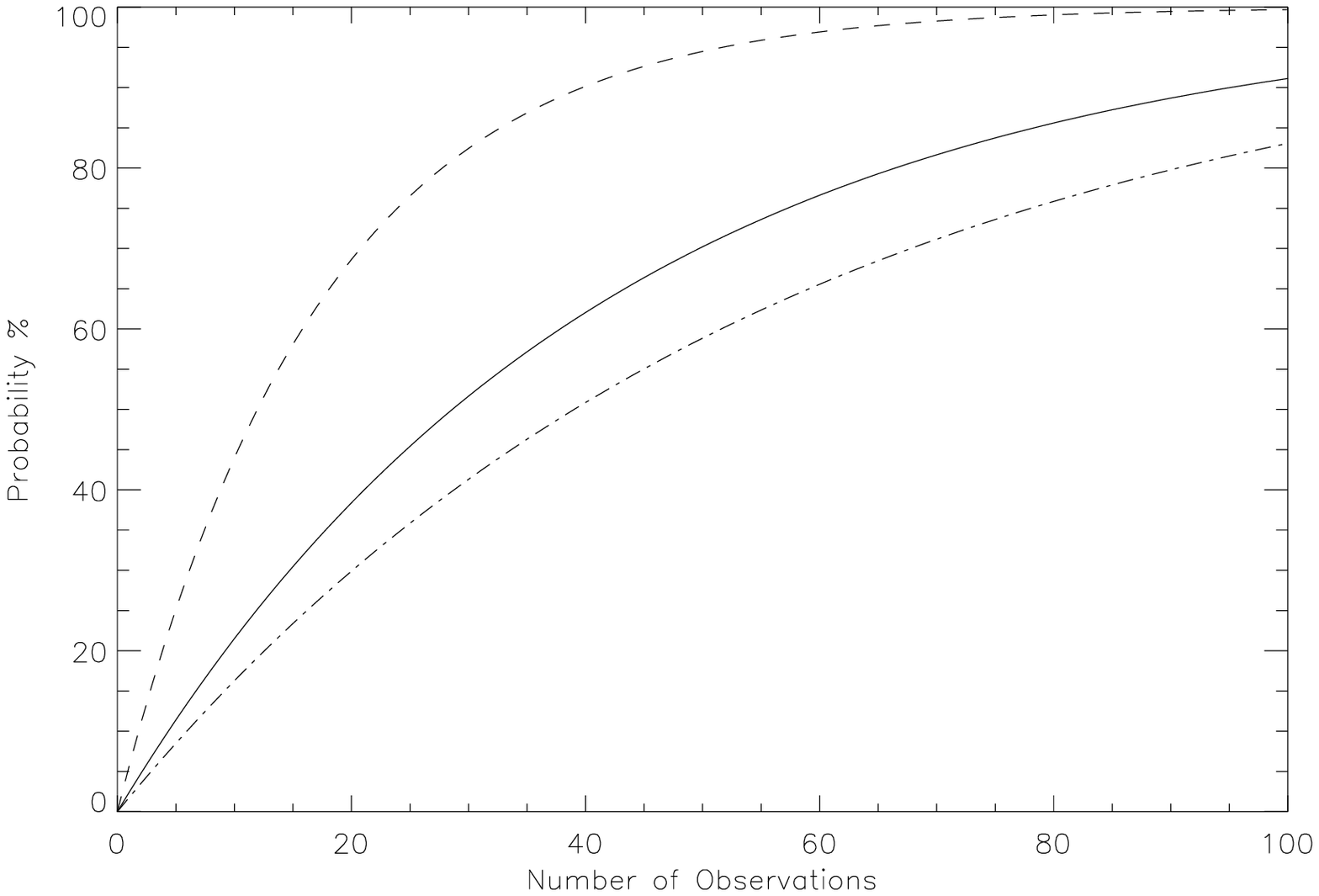} 
\caption{\label{plotseven} Probability of detection based on number of
observations and total number of comets per year=10 SP and 15
LP. Dashed line=maximum probability, solid line=average and
dotted/dashed line=minimum. (Left) Short period comets (Right) Long
period comets.}
\end{figure}

\begin{figure}
\includegraphics[scale=0.5]{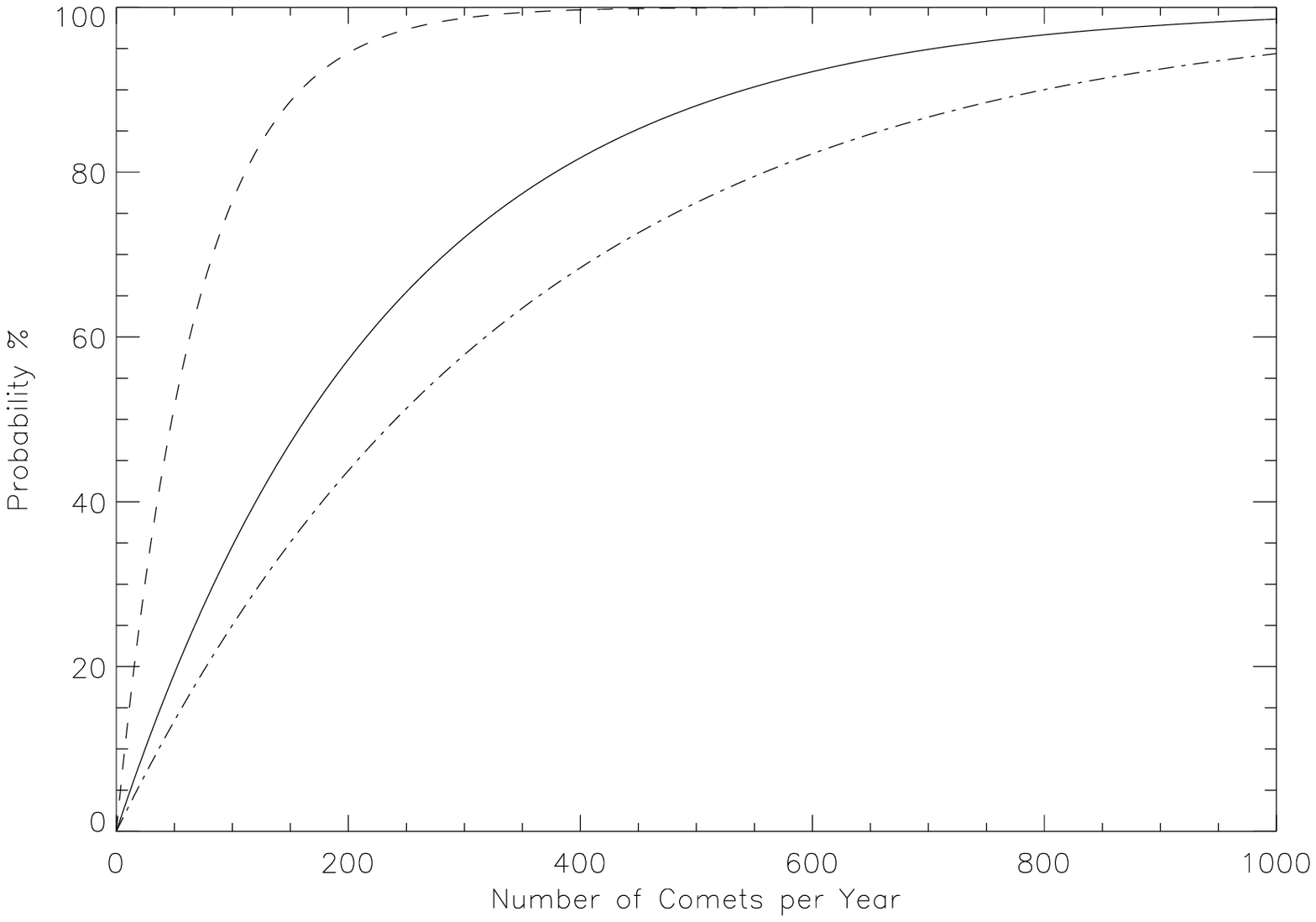}
\includegraphics[scale=0.5]{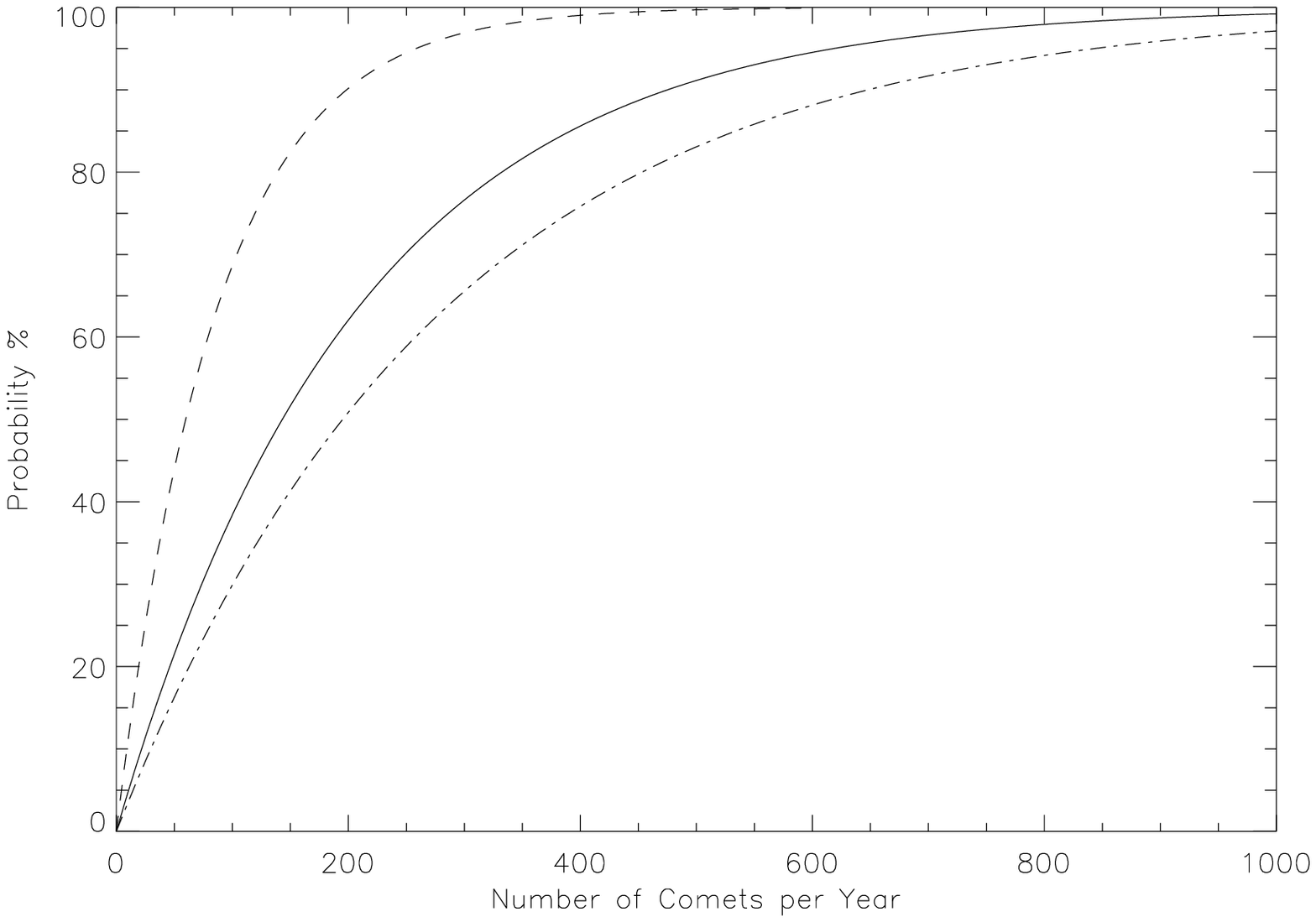} 
\caption{\label{ploteight} Probability of detection based on number of
comets per year and our total number of observations=6. Dashed
line=maximum probability, solid line=average and dotted/dashed
line=minimum. (Left) Short period comets. (Right) Long period comets.}
\end{figure}

The probability as a function of the number of observing sessions for
our long and short cometary models are plotted in Figure
\ref{plotseven}. The probability of detecting either type of periodic
comet from 6 observations (in two years, 2001-2003), ranges from 16\%
to 55\%, with an average probability of 20\%. According to these
probabilities then, we would need to make $\sim$20 observations to
achieve a 50\% likelihood of observing a comet. As a more conservative
estimate we only consider the scenario when we are looking directly
into the sodium tail, which would maximize the column density. This
then decreases D$_{sun}$ to essentially the stellar radius and gives a
probability of about half the values listed above and we would need to
make double the amount of observations to be certain of a 50\% chance.
The probability could also vary depending on the number of comets in
the system. In Figure \ref{ploteight} we show how the number of comets
affects the probability using the average, minimum and maximum time a
comet spends in front of the star and our current number of
observations. There would need to be $\sim$150 total comets per year
for us to have a 50\% chance with our current number of observations.

This implies that it would be necessary to either continue with more
observations of this system or expand our target list to cover many
more systems in order to have a better chance of detecting a comet
event.

\subsection{Theoretical Column Density}
For comparison, we have also conducted theoretical calculations on the
column density we would expect from typical comet events. Using
measurements of comet Hale-Bopp we calculate an approximate column
density of sodium in the tail. Using the production rate and lifetime
of neutral sodium atoms listed in \citet{Cremoneseal97} of
Q(Na{\footnotesize I})$ \simeq 5 \times 10^{25}$ atoms s$^{-1}$ and
$\tau=1.69 \times 10^{5}$ s at 1 AU (which for Hale-Bopp is very close
to perihelion, q=0.914), we estimate the column density of the tail to
be 4 $\times 10^{8}$ cm$^{-2}$. Soon after the discovery by
\citet{Cremoneseal97} of the sodium tail, \citet{Wilson98} discovered
that there was a second sodium tail (termed the diffuse sodium tail)
superimposed on the dust tail with a production rate of sodium very
similar to that determined for the pure sodium tail. If we consider
the sodium in both sodium tails and the contribution of sodium in
the coma, (which was estimated to be Q(Na{\footnotesize I})$ \simeq 1
\times 10^{25}$ atoms s$^{-1}$ by \citet{Combi97}), the total
production rate increases to 1 $\times 10^{26}$ atoms s$^{-1}$ and the
column density to $\sim 8\times 10^{8}$ cm$^{-2}$. \citet{Combi97}
argue that, based on solar abundances and the production rate of
oxygen (10$^{30}$ cm$^{-2}$), the total production rate of sodium
should be a few times 10$^{27}$ cm$^{-2}$, which would lead to a
column density of $\sim$8$\times 10^{9}$ cm$^{-2}$. Based on our upper
limits we would be able to detect a feature of this column density if
it were offset from the rest wavelength of the star. This is likely
since from Figure \ref{plottwo} it is shown that the majority of the
orbit will produce at least a 0.4 \AA\ shift. The upper limits
determined in the previous section are on the order of the column
density of sodium from a single periodic comet, which verifies that we
would be able to detect the passage of a single comet. With the
probability, this suggests an upper limit of $\sim$300 total comets
per year that come within 3 AU of HD209458 (based on a 95\%
probability(2$\sigma$)). Considering that the $\beta$ Pictoris system
has $\sim$300-1000 FEB's per year \citep{Lagrange92,Ferlet93}, this
upper limit is consistent with a more evolved system.

\section{Conclusion}
We monitored the planet-bearing solar-type star HD209458 for comet
activity using the HET over the course of 2 years with 6 observations
with S/N ratios between 70 and 258. Our observations of the sodium D1
line were examined for variable absorption components, as seen in
$\beta$ Pictoris and other stars, which are considered to be
signatures of this type of activity. From modelling, we determined
that a feature with a detectable column density of $\sim 10^{10}$
cm$^{-2}$ would have a FWHM of $\sim$0.05 \AA\ and a $W_{\lambda}$ of
$\sim$1 m\AA\ and could be shifted up to $\sim$0.5 \AA\ from the
photospheric line center. Simulations based on typical solar system
periodic comets showed that the probability of a detection from 6
observations was $\sim$20\% and was insensitive to the total elapsed
time between observations. The probabilities derived suggest that less
than $\sim$ 300 total comets per year come within 3 AU of HD209458 or
one would have likely been detected. No absorption features greater
than 3$\sigma$ due to cometary activity were identified. Our upper
limits for the sodium column density for each observation are
consistent with the column density expected from the sodium tail of
solar-type comets. Some possible reasons for the non-detection are 
\flushleft
\begin{enumerate}
\item The small number of observations. This system is expected to have
comet activity on a similar scale as the solar system since it is a
$\sim$5 Gyr old solar-type star \citep{Cody02}. With only 6
observations in 2 years, our chances of witnessing a comet crossing
were quite small. We plan to continue to monitor this star with many
more observations. 
\item A comet was passing but it was not detectable to us. The
inclination of the comets orbit might have been such that it was not
within our line of sight. Though, in the solar system, 87\% of short
period comets have inclinations within 30$^{\circ}$ and 81\% have
perihelia latitudes less than 10$^{\circ}$ from the ecliptic (all of
them are within 30$^{\circ}$) \citep{Biswas00}. Long period comet
inclinations are more scattered about the ecliptic, with only 47\%
having inclinations less than 90$^{\circ}$ and 63\% having perihelia
latitudes less than 30$^{\circ}$ \citep{Biswas00}. So, inclination may
factor into the probability when considering long period comets. 
\item The sodium feature was at the rest wavelength of the star. This
would occur if the comet was detected at perihelion, which would be a
less likely circumstance for periodic comets since they spend only a
small fraction of their time there. If it were the case though, based
on estimates of the column density of the sodium tail from a single
comet, our S/N in the core of the D1 line might not be sufficient to
uncover a signal. For our future observations we plan to obtain higher
S/N data. 
\end{enumerate}
Continued monitoring of HD209458 will occur over the next
year, and observations are planned for several other planet-bearing
stars. Through continued observations we can improve our upper limits
for the number of comets in the planetary system around HD209458 and
increase our chances of detecting a cometary event.

\acknowledgements 

We would like to thank John Debes for helping to initiate the
observations and for the use of his absorption profile code. We also
thank Robin Ciardullo for his helpful discussions concerning
probability. This work was supported by NASA grants NGA5-12115,
NAG5-11427, and NAG5-13320, NSF grants AST-0138235 and AST-0243090 and
the Penn State Eberly College of Science.

\begin{deluxetable} {ccccc}
\tabletypesize{\scriptsize}
\tablewidth{0pt}
\tablecaption{Observations \label{observations}}
\tablehead{ 
\colhead{Star} & 
\colhead{Observation Date} & 
\colhead{Heliocentric Velocity} & 
\colhead{Number of Exposures} & 
\colhead{Time per Exposure} \\
\colhead{} & 
\colhead{} & 
\colhead{km/s} & 
\colhead{} & 
\colhead{seconds}
} 
\startdata
HD209458 & 2001 Jul07 & 21.29 & 4 & 500 \\       
         & 2001 Jul09 & 20.78 & 2 & 900 \\       
         & 2002 Oct10 &-16.4  & 2 & 1500 \\ 
	 & 2002 Nov23 &-26.36 & 2 & 1500 \\ 
         & 2002 Dec09  &-26.46 & 2 & 1500 \\ 
	 & 2003 Jul14 &19.60  & 1 & 3600 \\
HD12235  & 28Aug02 & 23.60 & 2 & 1200 \\ 
	 & 31Aug02 & 22.70 & 3& 1000 \\ 
\enddata
\end{deluxetable}

\begin{deluxetable} {cccccccc}
\tabletypesize{\scriptsize}
\tablewidth{0pt}
\tablecaption{Stellar Parameters \label{params}}
\tablehead{
\colhead{Object} & 
\colhead{Age} & 
\colhead{Spectral Type\tablenotemark{a}} & 
\colhead{Radius\tablenotemark{b}} & 
\colhead{Teff\tablenotemark{b}} & 
\colhead{L\tablenotemark{b}} & 
\colhead{Vmag\tablenotemark{a}} & 
\colhead{Radial Velocity\tablenotemark{a}} \\
\colhead{} & 
\colhead{Gyr} & 
\colhead{} & 
\colhead{R$_{\odot}$} & 
\colhead{K} & 
\colhead{L$_{\odot}$} & 
\colhead{} & 
\colhead{km/s}
}
\startdata 
HD209458 & 5.2\tablenotemark{b}& GOV & 1.18  & 6000 & 1.61 & 7.7 & -14.8 \\
HD12235  & 5.39\tablenotemark{c}& G2IV &&&& 5.9  & -17.4 \\
\enddata
\tablenotetext{a}{Reference-SIMBAD data retrieval system (of the Astronomical Data center in Strasbourg, France)}
\tablenotetext{b}{Reference-\citet{Cody02}} 
\tablenotetext{c}{Reference-\citet{Ibukiyama02}}
\end{deluxetable}

\begin{deluxetable} {ccccc} 
\tabletypesize{\scriptsize}
\tablewidth{0pt}
\tablecaption{Sodium Column Density Upper Limits of
HD209458 System.\label{upplim}}
\tablehead{
\colhead{Observation Date} & 
\colhead{S/N in continuum} & 
\colhead{N(NaI) Upper Limit} & 
\colhead{S/N in line core} & 
\colhead{N(NaI) Upper Limit} \\
\colhead{} & 
\colhead{} & 
\colhead{cm$^{-2}$} & 
\colhead{} & 
\colhead{cm$^{-2}$}
}
\startdata
7July01 &  70 & 2.2$\times10^{10}$ & 23 & 6.5$\times10^{10}$ \\ 
9July01 & 110 & 1.4$\times10^{10}$ & 36 & 4.3$\times10^{10}$ \\ 
10Oct02 & 198 & 7.6$\times10^{9}$  & 67 & 2.3$\times10^{10}$ \\ 
23Nov02 & 237 & 6.4$\times10^{9}$  & 88 & 1.7$\times10^{10}$ \\ 
9Dec02  & 258 & 5.8$\times10^{9}$  & 88 & 1.7$\times10^{10}$ \\ 
14July03 & 204 & 7.4$\times10^{9}$ & 67 & 2.2$\times10^{10}$ \\
\enddata
\end{deluxetable}

line 
= 

\end{document}